\documentclass[aps,prc,fleqn,twocolumn,superscriptaddress,floatfix]{revtex4}

\usepackage{epsfig,graphicx}
\usepackage{amsmath,amssymb,amsfonts}

\newcommand{\Lav}{\langle L \rangle}
\newcommand{\Lavad}{\langle L^\dagger \rangle}

\begin{document}

\title{Center Domains and their Phenomenological Consequences}

\author{Masayuki Asakawa}
\affiliation{Department of Physics, Osaka University, Toyonaka 560-0043, Japan}

\author{Steffen A.~Bass}
\affiliation{Department of Physics, Duke University, Durham, NC 27708, USA} 

\author{Berndt M\"uller}
\affiliation{Department of Physics, Duke University, Durham, NC 27708, USA} 

\date{\today}

\begin{abstract}
We argue that the domain structure of deconfined QCD matter, which can be inferred from the properties of the Polyakov loop, can simultaneously explain the two most prominent experimentally verified features
of the quark-gluon plasma, namely its large opacity as well as its near ideal fluid properties.
\end{abstract}

\maketitle

One of the major achievements of the experimental program at the Relativistic Heavy Ion Collider (RHIC) is the creation of a novel state of hot and dense QCD matter dubbed the strongly interacting Quark-Gluon Plasma (sQGP) \cite{Arsene:2004fa,Adcox:2004mh,Back:2004je,Adams:2005dq,Gyulassy:2004zy,Muller:2006ee}. The characterization of the properties of the sQGP is based on three major discoveries: (1) the measurement of strong elliptic flow and the success of relativistic viscous hydrodynamics with a very small value of the shear viscosity $\eta$ to entropy density $s$ ratio close to the conjectured quantum lower bound \cite{Kovtun:2004de} that argues for the near perfect liquid nature of the sQGP and a very short thermalization time of less than 1 fm/c, (2) the measurement of very strong suppression of high-momentum particles and rapid redistribution of the jet energy into the whole solid angle, which  is indicative of the large opacity of the produced matter, and (3) the observed constituent quark number scaling law for the elliptic flow of identified hadrons as predicted by the parton recombination model  \cite{Fries:2003kq,Greco:2003mm,Molnar:2003ff,Fries:2008hs} that provides the most direct evidence for the formation of hadrons from a deconfined system of interacting patrons. All three discoveries have by now been confirmed in measurements of Pb+Pb collisions at the CERN Large Hadron Collider (LHC) \cite{Muller:2012zq}.

In this work we shall focus on the dynamical properties of the sQGP and shall present a picture in which
the low specific viscosity and high color opacity  can be understood in a consistent fashion.
 Most calculations of the jet energy-loss are based on perturbative QCD and are only sensitive to the gluon content of the matter, but do not distinguish between different microscopic structures related to quark confinement or chiral symmetry breaking. Lattice calculations are not yet able to give reliable results for transport properties of the sQGP that can provide clues about its structure. Holographic approaches to strongly coupled supersymmetric gauge theories have been used to model dynamical properties of the sQGP, but it remains unclear how well these gauge plasmas mirror the physics of the QGP.

So far, very little attention has been paid to the non-perturbative structure of the gauge field configurations in the quark-gluon plasma and how it may affect the features observed in the experiments at RHIC and LHC. In particular, our focus here is on the Polyakov loop. We shall argue that the sQGP produced in relativistic heavy ion collisions has a domain structure based on the different minima of the Polyakov loop potential in the deconfined phase, and that this domain structure can be instrumental for generating the large opacity and small value of $\eta/s$ seen in the sQGP.

Two characteristic features distinguish the quark-gluon plasma from the hadron phase: restored chiral symmetry in the light quark sector and color deconfinement. The order parameter which characterizes the confinement-deconfinement transition in the pure SU(3) gauge theory is the Polyakov loop $L$. It is defined as\begin{equation}
L(\vec{x})=\frac{1}{3} {\rm tr~P}\exp
\left [ ig \int_{0}^{1/T}A_4(\tau,\vec{x})d\tau \right ],
\label{eq:L(x)}
\end{equation}
where P is an ordering operator with regard to the imaginary time $\tau$, $A_4 = iA^0 = iA^{0a}\lambda^{a}/2$ with the Gell-Mann matrices $\lambda^a$ ($a=1,\cdots,8$), $T$ the temperature, and $\vec{x}$ the three-dimensional spatial position. In the pure gauge theory, the Polyakov loop and the free energy of a heavy static quark, $F_Q(T)$, are related by
\begin{equation}
F_Q(\vec{x},T) = -T\ln | \langle L(\vec{x},~T) \rangle | ,
\label{eq:FQ}
\end{equation}
where $\langle\cdot\rangle$ stands for the thermal average. $\langle L(\vec{x},~T) \rangle$ vanishes in the confined phase.  In the deconfined phase it takes the value of one of the three elements of $Z_3$, the center of SU(3): $\exp(i2\nu\pi/3)$ with $\nu=0,1,2$. Thus, the confined phase is $Z_3$ symmetric, while the $Z_3$ symmetry is spontaneously broken in the hot deconfined phase. Lattice calculations in the deconfined phase are usually carried out around one of these three states. Here we explore how the center symmetry affects the properties of the hot matter created in relativistic heavy ion collisions. 

The fundamental Polyakov loop (\ref{eq:L(x)}), $L^{(3)} \equiv L$, governs the interaction of a static quark with the thermal gauge field. The interaction of particles carrying color charge in the adjoint representation, such as gluons, is described by the adjoint Polyakov loop. There is good evidence from lattice simulations \cite{Gupta:2006qm,Mykkanen:2012ri} that the expectation value of the adjoint Polyakov loop $L^{(8)}$ is given by a power of $\langle L^{(3)} \rangle$ equal to the ratio of the Casimir operators of the adjoint and fundamental representation of SU(3): $\langle L^{(8)} \rangle \approx \langle L^{(3)} \rangle^{9/4}$. This means that gluons interact even more strongly with the center domain walls than quarks. 

It is now widely believed that the quark-gluon plasma is formed through a non-equilibrated precursor state called {\em glasma} \cite{Lappi:2006fp}. The glasma is a longitudinally slowly varying, classical gauge field configuration whose structure in the transverse direction is characterized by the saturation scale $Q_s$, with $Q_s \approx 1.5$ GeV at RHIC and $Q_s \approx 2$ GeV at LHC \cite{Albacete:2011fw}. 
Through thermalization this gauge field materializes into the quark-gluon plasma. The values of Polyakov loop in the plasma will locally cluster around the one of the preferred values $\exp(i2\nu\pi/3)$ where the potential has a minimum. Since the original transverse correlation length of the gauge fields is of order $Q_s^{-1}$, causality dictates that it can be at most of order $\tau_{\rm th}$ at the moment of thermalization.  In other words, the distribution of Polyakov loop values must assume a domain structure on the transverse plane with a typical transverse domain size $R_d$ constrained by $Q_s^{-1} < R_d < \tau_{\rm th}$. Each domain is separated from other domains by potential walls, where the Polyakov loop values interpolate between the different $Z_3$ values. We call these domains {\em center domains} \cite{Pisarski:2000eq,Dumitru:2000in}. We note here that the existence of center domains in the thermal quark-gluon plasma phase was recently demonstrated on the lattice even in the presence of dynamical quarks with the physical masses \cite{Borsanyi:2010cw,Danzer:2010ge}.

\begin{figure}[htb]   
\includegraphics[width=0.6\linewidth]{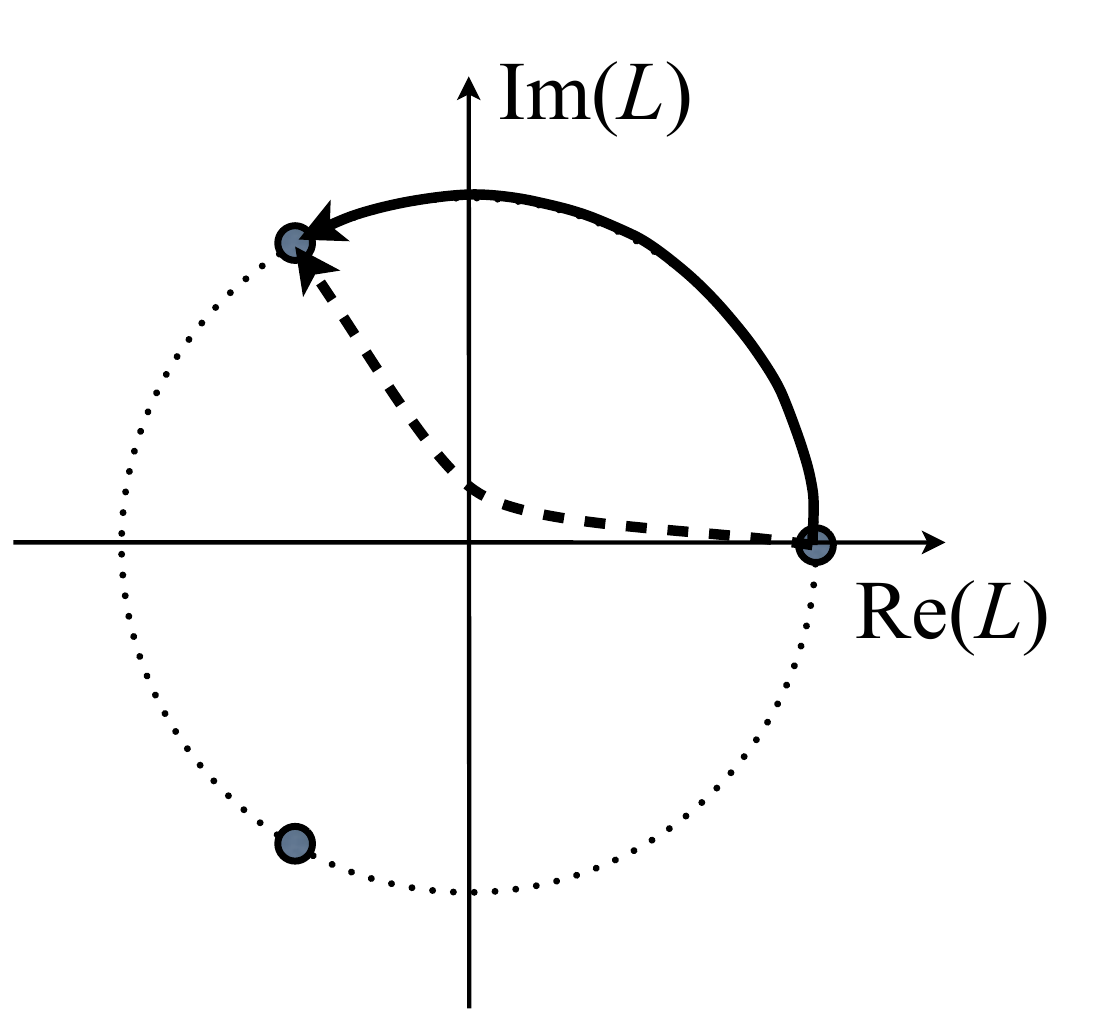}
\caption{In the deconfined phase of QCD, center domains are characterized by one of the three values $\exp(i2\nu\pi/3)$ of the expectation value of the Polyakov loop $\Lav$ corresponding to the minima of the Polyakov loop potential. In the boundary wall between two domains, $\Lav$ changes from one minimum to another. The figure shows two types of schematic trajectories.}
\label{fig:path}
\end{figure}

\begin{figure}[htb]   
\includegraphics[width=0.95\linewidth]{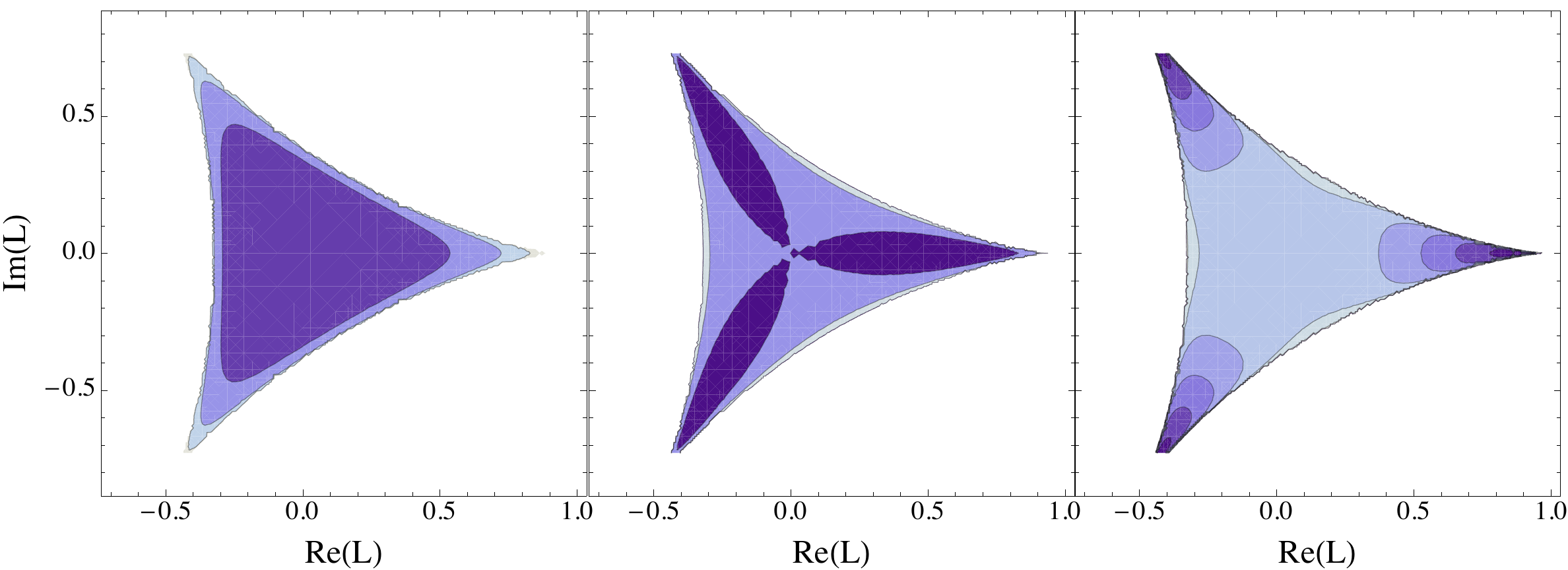}
\caption{Contour plots of the SU(3) Polyakov loop potential (\ref{eq:UL}) at $T = 100$ MeV (left), $T = 300$ MeV (middle), and $T = 500$ MeV (right). The transition from the confined to the deconfined phase is clearly visible by a shift of the minimum from $\Lav = 0$ to $\Lav = \exp(i2\nu\pi/3)$.}
\label{fig:PolPot}
\end{figure}

There are two possible types of trajectories for the Polyakov loop between two $Z_3$ minima. Either the expectation value of the Polyakov loop remains of unit modulus, but its phase changes gradually by $2\pi/3$ inside the domain wall (solid path in Fig.~\ref{fig:path}), or the modulus of the  Polyakov loop becomes smaller than unity and abruptly changes its phase near the origin (dashed path in Fig~\ref{fig:path}).  In order to determine which case is realized, we need the information of the Polyakov loop potential in the deconfined phase. Its form has been empirically determined by a combination of analytical considerations and by fitting the lattice results for the pressure as a function of temperature. A commonly used form in the so-called Polyakov loop--Nambu--Jona-Lasinio (PNJL) model is \cite{Fukushima:2003fw}
\begin{equation}
U(\Lav) = - b T \left[ 54 e^{-a/T} |\Lav|^2 + \ln P(\Lav,\Lavad) \right] .
\label{eq:UL}
\end{equation}
with $P(z,\bar{z})=1 - 6 |z|^2 - 3|z|^4 + 4(z^3 + \bar{z}^3)$ arising from the SU(3) Haar measure and numerical constants $a = 0.664~{\rm GeV}$ and $b = 0.0075~{\rm GeV}^3$. $T$ denotes the temperature. Contour plots of the Polyakov loop potential (\ref{eq:UL}) at three different temperatures below, near and above the deconfinement temperature (100, 300, and 500 MeV) are shown in Fig.~\ref{fig:PolPot}.  Since the potential $U(\Lav)$ is deduced from lattice simulations, its empirical form is most reliable in the vicinity of the potential minima, i.e. $\Lav=\exp(i2\nu\pi/3)$, but it is not well known far away from these minima. The Polyakov loop potential is thus not sufficiently accurately determined to reliably answer the question with certainty, which trajectory $\Lav$ takes inside the domain wall.

Assuming the widely adopted potential form (\ref{eq:UL}), the Polyakov loop prefers to take values around zero inside the domain walls between deconfined regions, because trajectories corresponding to the dashed path in Fig.~\ref{fig:path} involve the lowest potential barrier. Since the expectation value of the Polyakov loop in the confined phase vanishes, center domains are thus separated from neighboring ones by walls that are characterized by gauge field configurations similar to those in the confined phase of the gauge theory. This insight forms the basis of our following discussion.

We now proceed to consider the possible phenomenological consequences of the presence of the center domains in the quark-gluon plasma produced in heavy ion collisions. First, the walls act as a potential barrier for partons with momentum smaller than the confining scale. This can be seen by invoking the following argument. The free energy of a single static quark is related to the expectation value of the Polyakov loop by the well known expression (\ref{eq:FQ}). A vanishing average Polyakov loop thus corresponds to an infinite energy of an isolated heavy quark. For light quarks, this argument does not imply confinement because of the possibility of pair creation, but it implies that it is energetically unfavorable for any quark to propagate through a region with $\Lav\approx 0$.

Thus, partons with thermal momenta cannot classically penetrate the walls but instead are reflected on them. Consequently, the mean free path $\lambda_{\rm f}$ of partons becomes of order $R_d/2$, where $R_d$ is the average domain size. Inserting this result into the kinetic theory formula for the shear viscosity,
\begin{equation}
\eta = \frac{1}{3}n\bar{p}\lambda_{\rm f},
\end{equation}
where $n$ denotes the particle density and $\bar{p}$ is the thermal momentum, one obtains
\begin{equation}
\frac{\eta}{s} \approx \frac{1}{8} T R_d ,
\end{equation}
with $\bar{p}\approx 3T$, and $s/n\approx 4$ valid for massless particles. For $T=400~$MeV and $R_d=0.5~$fm
this results in a value of $\eta/s \approx 0.125$.
This mechanism of lowering the shear viscosity resembles that of the anomalous viscosity \cite{Asakawa:2006tc,Asakawa:2006jn}, but the mechanism responsible for a small mean free path is different.

For patrons with higher momenta, which can cross the boundary between domains, the system of domain walls acts like the combination of a frequency collimator and an irregular undulator. Because they serve as a reflective barrier to soft gluons, the domain walls constitute a very effective mechanism for the ``frequency collimation''  effect \cite{CasalderreySolana:2010eh}, which has been conjectured to be responsible for the large di-jet asymmetries observed in Pb+Pb collisions at the LHC \cite{Aad:2010bu,Chatrchyan:2011sx}.  The stripped soft gluon component of the nascent jet is quickly restored, because the walls alternately decelerate and then re-accelerate fast moving partons. This can be seen by expanding the expression of the average of the Polyakov loop and making a Gaussian approximation:
\begin{eqnarray}
\langle L(\vec{x}) \rangle
& = & \frac{1}{3} \left\langle {\rm tr~P}\exp
\left [ ig \int_{0}^{1/T}A_4(\tau,\vec{x})d\tau \right ] \right \rangle 
\nonumber \\
& \approx &
\exp \left [ -\frac{g^2}{2T^2} {\rm tr} \langle A^0(\vec{x})^2  \rangle \right ] .
\end{eqnarray}
Inside the domain walls, where the Polyakov loop approximately vanishes, the gauge field $A^0$ fluctuates with large amplitude, and partons thus feel sudden changes of the zeroth component of the external vector potential through the coupling in the covariant derivative. This leads to strong and uncorrelated radiation of gluons every time a parton crosses one of the walls, which are distributed randomly with the density of the order of $R_d^{-3}$. 

The center domains thus provide a novel, nonperturbative mechanism for jet quenching, which is different from other mechanisms in two aspects.  First, it distinguishes between the QCD phases: jet quenching by center domains is caused by the confining phase composing the domain walls. In the conventional mechanisms, jet quenching is not affected by which phase the matter is in, but only by how large the gluon density is. Second, the gluon radiation is very different at low and high frequencies: At frequencies higher than the height of the domain wall barrier, $\omega > \omega_c$, gluons propagate as quasi-particles, but radiation of gluons is enhanced by partially coherent emission from the crossing of many domain walls. At low frequencies, radiated gluons are strongly scattered by the domain walls and propagate diffusively, not ballistically. 

To obtain an estimate of the energy loss in a single wall crossing we simply assume that all gluons below a critical frequency $\omega_c$ are scattered away from the parent parton. The energy loss is then given by
\begin{equation}
\Delta E = \int_0^{\omega_c} \omega \frac{dN_g}{d\omega} d\omega ,
\end{equation}
where $dN_g/d\omega$ denotes the frequency spectrum of gluons accompanying the hard parton.  If just the  Li\'enart-Wiechert field of the hard parton is restored after a wall crossing, the gluon spectrum is given in the Weizs\"acker-Williams approximation by:
\begin{equation}
\frac{dN_g}{d\omega} \approx \frac{C_2 \alpha_s}{\pi\omega} \ln(\omega/\omega_0)\, \theta(\omega-\omega_0) ,
\end{equation}
where we have introduced an infrared cut-off $\omega_0 \sim O(T)$ accounting for the finite size of the domains, which limits the ability of the parton to radiate very soft gluons.  $C_2$ denotes the Casimir for the color charge of the hard parton. The energy loss per unit length is then
\begin{equation}
\frac{dE}{dx} \approx \frac{C_2 \alpha_s}{\pi R_d} \{\omega_c\ln(\omega_c/\omega_0)
-(\omega_c-\omega_0)\}
\end{equation}
Estimating the cut-off frequency as $\omega_c \approx 1-2$ GeV, $\omega_0 \approx 0.4$ GeV, and $R_d \approx 0.5$ fm, we find values of $dE/dx$ in the range $(0.2-1) \, C_2 \, \alpha_s$ GeV/fm. 

In addition to this energy loss associated with energy collimation, for gluon frequencies above $\omega_c$, radiation is enhanced by the interaction of the energetic parton with several domain walls. Making use of the undulator analogy, the characteristic frequency of radiation emitted by a massive parton traversing the system of domain walls is given by \cite{Brau:1990aha}
\begin{equation}
\omega_L = 4\pi \gamma^2/R_d ,
\end{equation}
where $\gamma = E/m$ is the Lorentz factor of the parton. Obviously, these nonperturbative effects can lead to enormous energy loss of moderately heavy quarks, such as charm quarks, at energies a few times their rest mass.  

\begin{figure}[htb]   
\includegraphics[width=0.9\linewidth]{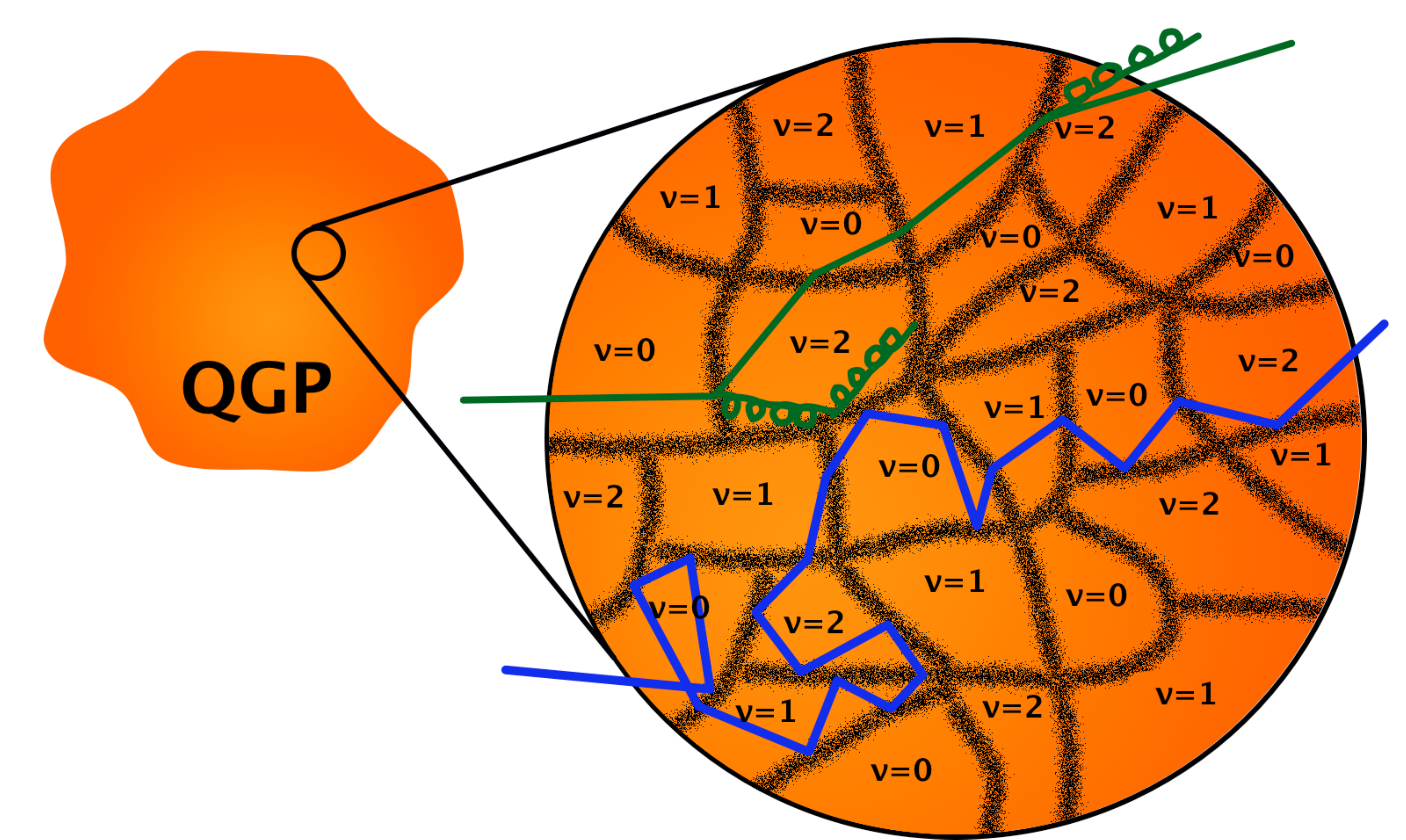}
\caption{Schematic representation of the domain structure of a small volume of QGP matter
with two patrons traversing it. A fast parton crossing the walls of a center domain will
radiate (soft) gluons that may reflect from the domain walls, leading to large energy loss
and rapid isotropization of the jet energy. Soft partons may reflect multiple times off
domain walls while propagating through the QGP.}
\label{fig:cartoon}
\end{figure}

Another consequence of the center domain scenario is that the momenta of the emitted soft gluons are immediately randomized by the reflection on the walls. This property can explain the almost complete redistribution of the quenched energy over the whole solid angle observed in Pb+Pb collisions at LHC \cite{Tonjes:2011zz}.  Figure~\ref{fig:cartoon} depicts two partons  propagating through a small volume of QGP matter: a hard parton radiates (soft) gluons while crossing domain walls whereas soft partons may reflect off the domain walls.

The time evolution of the center domains has been studied in a different context \cite{Gupta:2010pp}. Gupta {\em et al.} assumed that center domains are created by the tunneling of the Polyakov loop expectation values in the early stage of the quark-gluon plasma formation. The expected size of the domains is then much larger than $Q_s^{-1}$, and it is assumed to possess the boost-invariant structure for simplicity. According to their numerical simulation, domains merge if their initial values of $\Lav$ lie in the same minimum, while domains do not merge if their initial values of $\Lav$ correspond to different minima. 

Applying the results of ref.~\cite{Gupta:2010pp} to our scenario, we conclude that both the size and number of the initially created domains does not change much during the time evolution of the quark-gluon plasma. The mean free path of partons does not increase substantially beyond the range of $Q_s^{-1}$ estimated above, and the correlation length of the gauge fields is limited by this scale, while the system is strongly interacting. This implies that hydrodynamics remains applicable throughout the expansion until the deconfinement  transition.

When the temperature approaches the pseudo-critical temperature $T_c$, the potential difference between $\Lav=0$ and $\Lav = \exp(i2\nu\pi/3)$ becomes smaller, the walls become wider, and the deconfined domains become smaller. Eventually, deconfined domains fragment into individual hadrons, and the walls become the nonperturbative QCD vacuum. The center domains thus describe the evolution of the quark-gluon plasma consistently from its birth to hadronization and help to explain the strongly coupled nature of the quark-gluon plasma including all the observed properties of the quark-gluon plasma from its hydrodynamical behavior to jet quenching.

In summary, we have argued that the center domains are an important facet of the evolution of the quark-gluon plasma from its birth up to hadronization. They naturally explain the strongly coupled nature of the quark-gluon plasma including its major observed properties from its nearly ideal hydrodynamical behavior to strong jet quenching.

B.~M. and S.~A.~B. acknowledge support by  U.S. Department of Energy grants DE-FG02-05ER41367 and DE-SC0005396. M.~A. was supported by Grants-in-Aid for Scientific Research 23540307 from JSPS of Japan.


\end{document}